\documentclass[12pt]{article}
\usepackage{axodraw,bbold}

\catcode`@=12
\begin{document}
\thispagestyle{empty}
\begin{flushright} 
UCRHEP-T452\\ 
June 2008\
\end{flushright}
\vspace{0.5in}
\begin{center}
{\LARGE	\bf Axionic Extensions of the\\ Supersymmetric Standard Model\\}
\vspace{1.0in}
{\bf Ernest Ma\\}
\vspace{0.2in}
{\sl Department of Physics and Astronomy, University of California,\\}
\vspace{0.1in}
{\sl Riverside, California 92521, USA\\}
\vspace{1.5in}
\end{center}

\begin{abstract}\
The Supersymmetric Standard Model is a benchmark theoretical framework for 
particle physics, yet it suffers from a number of deficiencies, chief among 
which is the strong CP problem.  Solving this with an axion in the context 
of selected new particles, it is shown in three examples that other problems 
go away \emph{automatically} as well, resulting in $(-)^L$ and $(-)^{3B}$ 
conservation, viable combination of two dark-matter candidates, successful 
baryogenesis, seesaw neutrino masses, and verifiable experimental consequences 
at the TeV energy scale.
\end{abstract}

\newpage
\baselineskip 24pt

A benchmark theoretical framework for the study of particle physics is the 
Supersymmetric Standard Model.  It solves the hierarchy problem of scalar 
masses, yet it suffers from a number of deficiencies.  Consider its 
particle content as shown in Table 1.

\begin{table}[htb]
\caption{Particle content of the Supersymmetric Standard Model.}
\begin{center}
\begin{tabular}{|c|c|c|c|}
\hline 
Superfield & $SU(3)_C \times SU(2)_L \times U(1)_Y$ & $(-)^L$ & $(-)^{3B}$ \\ 
\hline
$Q \equiv (u,d)$ & $(3,2,1/6)$ & + & --  \\ 
$u^c$ & $(3^*,1,-2/3)$ & + & --  \\ 
$d^c$ & $(3^*,1,1/3)$ & + & --   \\ 
\hline
$L \equiv (\nu,e)$ & $(1,2,-1/2)$ & -- & + \\ 
$e^c$ & $(1,1,1)$ & -- & +  \\ 
\hline
$\Phi_1 \equiv (\phi^0_1,\phi^-_1)$ & $(1,2,-1/2)$ & + & +  \\ 
$\Phi_2 \equiv (\phi^+_2,\phi^0_2)$ & $(1,2,1/2)$ & + & + \\ 
\hline
\end{tabular}
\end{center}
\end{table}

Without the imposition of $(-)^L$, the terms $L \Phi_2$, $L L e^c$, and 
$L Q d^c$ would be allowed.  Without the imposition of $(-)^{3B}$, the 
term $u^c d^c d^c$ would be allowed.  To prevent rapid proton decay, 
either $(-)^L$ or $(-)^{3B}$ or both must be imposed.  If both are enforced, 
$R$ parity is conserved and the lightest supersymmetric particle (LSP) 
is a good dark-matter candidate.  This is called the Minimal Supersymmetric 
Standard Model (MSSM), and because of its minimal particle content, it 
also conserves additive lepton number $L$ and additive baryon number $B$, 
except for nonperturbative sphaleron effects which violate $B+L$ but conserve 
$B-L$.  Another shortcoming of the MSSM is the appearance of the allowed 
$\mu \Phi_1 \Phi_2$ term. Since $\mu$ is unrelated to any symmetry breaking, 
there is no guarantee that it is of order the supersymmetry breaking scale, 
which has to be itself close to the electroweak breaking scale, for a 
successsful phenomenological description of all particle interactions.  
Further, neutrino masses are absent and the strong CP problem is unresolved 
as in the (nonsupersymmetric) Standard Model.  In the following, it will be 
shown in three examples how an axionic extension with selected new 
particles will do away with all these deficiencies.

The strong CP problem is the appearance of the instanton-induced term 
\cite{cdg76,jr76}
\begin{equation}
{\cal L}_\theta = \theta_{QCD} {g_s^2 \over 64 \pi^2} \epsilon_{\mu \nu \alpha 
\beta} G_a^{\mu \nu} G_a^{\alpha \beta}
\end{equation}
in the effective Lagrangian of quantum chromodynamics (QCD), where $g_s$ is 
the strong coupling constant, and
\begin{equation}
G_a^{\mu \nu} = \partial^\mu G_a^\nu - \partial^\nu G_a^\mu + g_s f_{abc} G_b^\mu 
G_c^\nu
\end{equation}
is the gluonic field strength.  This term is odd under CP and if 
$\theta_{QCD}$ is of order unity, the neutron electric dipole moment would be 
$10^{10}$ times its present experimental upper limit ($0.63 \times 10^{-25} e$ 
cm) \cite{h99}.  This undesirable situation is most elegantly resolved by 
invoking a dynamical mechanism \cite{pq77} to relax the above $\theta_{QCD}$  
parameter (including all contributions from colored fermions) to zero.  
However, this requires an anomalous global $U(1)_{PQ}$ symmetry which is 
broken at the scale $f_a$ and results necessarily \cite{we78,wi78} in a 
very light pseudoscalar particle called the axion, which has not yet been 
observed \cite{rb00}.

To reconcile the nonobservation of an axion in present experiments and the 
constraint $10^9$ GeV $< f_a < 10^{12}$ GeV from astrophysics and cosmology 
\cite{r99}, three types of ``invisible'' axions have been discussed. (I) 
The DFSZ solution \cite{dfs81,z80} introduces a heavy singlet scalar field 
as the source of the axion but its mixing with the doublet scalar fields 
(which couple to the usual quarks) is very much suppressed. (II) The KSVZ 
solution \cite{k79,svz80} also has a heavy singlet scalar field but it 
couples only to new heavy colored fermions. (III) The gluino solution 
\cite{dm00} identifies the $U(1)_R$ of superfield transformations with 
$U(1)_{PQ}$ so that the axion is a dynamical phase attached to the gluino 
(which contributes to $\theta_{QCD}$ because it is a colored fermion) 
as well as all other superparticles.  

In a supersymmetric extension of the Standard Model, it is also important 
that the breaking of $U(1)_{PQ}$ at the large scale $f_a$ does not break the 
supersymmetry as well.  This may be accomplished using three (or more) singlet 
superfields in various ways, for the gluino solution \cite{dms00,dm01,m07}, 
the DFSZ solution \cite{m01,m08_1}, and a combination of the KSVZ and DFSZ 
solutions \cite{m03}.  The identification of $f_a$ as 
the seesaw scale of neutrino mass generation may also be achieved 
\cite{dms00,m07,m01,m08_1,m03}. 

To allow $\nu$ to acquire a mass, the conventional method is to add a 
neutral singlet fermion $N$, so that the terms
\begin{equation}
f_N \Phi_2 L N - {1 \over 2} m_N N N + H.c.
\end{equation}
may be added to the Lagrangian of the Standard Model (SM).  If $L = -1$ 
for $N$ is imposed so that $m_N=0$, then $\nu$ pairs up with $N$ to 
form a Dirac fermion of mass $f_N \langle \phi^0_2 \rangle$.  If $m_N \neq 0$ 
is allowed and $m_N >> f_N \langle \phi^0_2 \rangle$ is assumed, then the 
small Majorana mass $m_\nu \simeq - f_N^2 \langle \phi^0_2 \rangle^2/m_N$ 
is obtained, realizing the famous canonical seesaw mechanism.  However, 
there are actually three (and only three) tree-level realizations \cite{m98} 
of the unique dimension-five operator \cite{w79}
\begin{equation}
{\cal L}_5 = - {f_{ij} \over 2 \Lambda} (L_i \Phi)(L_j \Phi)
\end{equation}
for Majorana neutrino mass in the SM, the second utilizing a heavy scalar 
triplet $(\xi^{++},\xi^+,\xi^0)$ and the third a heavy fermion triplet 
$(\Sigma^+,\Sigma^0,\Sigma^-)$ per family.  These latter options will be 
used in the three examples to follow, because the singlet $N$ will be 
considered instead as a fermion odd under $(-)^{3B}$ \cite{m88,m08_2}.
The goal is to find an extension of the SM such that the proper $U(1)_{PQ}$ 
assignment will result in $(-)^L$ and $(-)^{3B}$ conservation automatically, 
together with other desirable consequences.

\begin{table}[htb]
\caption{Particle content of Example 1.}
\begin{center}
\begin{tabular}{|c|c|c|c|c|}
\hline 
Superfield & $SU(3)_C \times SU(2)_L \times U(1)_Y$ & $U(1)_{PQ}$ & 
$(-)^L$ & $(-)^{3B}$ \\ 
\hline
$Q \equiv (u,d)$ & $(3,2,1/6)$ & 1/2 & + & -- \\ 
$u^c$ & $(3^*,1,-2/3)$ & 1/2 & + & -- \\ 
$d^c$ & $(3^*,1,1/3)$ & 1/2 & + & -- \\ 
$N$ & $(1,1,0)$ & 1/2 & + & -- \\
\hline
$L \equiv (\nu,e)$ & $(1,2,-1/2)$ & 0 & -- & + \\ 
$e^c$ & $(1,1,1)$ & 1 & -- & + \\ 
\hline
$\Phi_1 \equiv (\phi^0_1,\phi^-_1)$ & $(1,2,-1/2)$ & --1 & + & + \\ 
$\Phi_2 \equiv (\phi^+_2,\phi^0_2)$ & $(1,2,1/2)$ & --1 & + & + \\ 
\hline
$h$ & $(3,1,-1/3)$ & --1 & + & + \\ 
$h^c$ & $(3^*,1,1/3)$ & --1 & + & + \\ 
\hline
$(\xi^{++}_1,\xi^+_1,\xi^0_1)$ & (1,3,1) & 0 & + & + \\
$(\xi^0_2,\xi^-_2,\xi^{--}_2)$ & (1,3,--1) & 2 & + & + \\
\hline
$S_2$ & $(1,1,0)$ & 2 & + & + \\ 
$S_1$ & $(1,1,0)$ & --1 & + & + \\ 
$S_0$ & $(1,1,0)$ & --2 & + & + \\ 
\hline
\end{tabular}
\end{center}
\end{table}

\newpage
\noindent {\bf (I)} As a first example, consider the scalar-triplet mechanism 
of neutrino mass  in a supersymmetric context. In Table 2, the PQ charges of 
the superfields of this construction and their \emph{derived} $(-)^L$ and 
$(-)^{3B}$ values are listed.  The complete superpotential is given by
\begin{eqnarray}
W_1 &=& m_0 S_0 S_2 + \lambda_1 S_1 S_1 S_2 + \lambda_2 S_1 N N 
+ \lambda_3 S_0 \xi_1 \xi_2 + f_1 S_2 \Phi_1 \Phi_2 + f_2 S_2 h h^c 
+ f_3 Q Q h \nonumber \\ 
&+& f_4 u^c d^c h^c + f_5 h d^c N + f_d \Phi_1 Q d^c 
+ f_u \Phi_2 Q u^c + f_e \Phi_1 L e^c + f_L L L \xi_1 
+ f_\phi \Phi_2 \Phi_2 \xi_2.
\end{eqnarray}
It is easy to see that $(-)^L$ and $(-)^{3B}$ are automatically conserved 
in this case. In contrast, the particle content of the model proposed in 
Ref.~\cite{m08_1} requires either $(-)^L$ or $(-)^{3B}$ to be imposed in 
addition to $U(1)_{PQ}$. Note that the only allowed mass term is $m_0$ which 
is thus expected to be large.  With $W_1$ of Eq.~(5), it has been shown 
\cite{m01,m08_1} that it is possible to break $U(1)_{PQ}$ spontaneously at 
the scale $m_0$ without breaking the supersymmetry.  The soft breaking of 
supersymmetry will then introduce another (much smaller) scale $M_{SUSY}$, 
with the result $u_1 = \langle S_1 \rangle$ and $u_0 = \langle S_0 \rangle$ 
are of order $m_0$, whereas $u_2 = \langle S_2 \rangle$ is of order $M_{SUSY}$. 
This means that the so-called $\mu$ problem in the MSSM is solved because 
$\mu = f_1 u_2$.  Similarly, the exotic $h$ quark has the mass $f_2 u_2$ 
and should be observable at the Large Hadron Collider (LHC).  As for the 
masses of $N$ and $\xi_{1,2}$, they are given by $2 \lambda_2 u_1$ and 
$\lambda_3 u_0$ respectively, with the axion contained in the dynamical 
phase of $(u_1 S_1 + 2 u_0 S_0)/\sqrt{u_1^2 + 4 u_0^2}$.  For the details 
of how supersymmetry remains unbroken at the axion scale, see 
Ref.~\cite{m08_1}. 

The scale $m_0$ determines the axion scale as well as $m_N$ and $m_\xi$. 
A baryon asymmetry is generated \cite{m08_2} in the early Universe by the 
decay of the lightest $N$ into $h \bar{d}$ and $\bar{h} d$ with the 
subsequent decay of $h$ into $\bar{u} \bar{d}$ and $\bar{h}$ into $u d$.  
Below the scale $m_N$, additive baryon number is conserved, hence the 
intervention of sphalerons which violate $B+L$ but not $B-L$ during the 
electroweak phase transition will allow a baryon asymmetry to remain as 
observed, in analogy to the usual leptogenesis scenario \cite{dnn08}.
The decay of the lightest $\xi_{1,2}$ may also generate a lepton asymmetry 
\cite{ms98,hms01}, but it is not necessary here so that no more than one 
pair of $\xi_{1,2}$ superfields is needed. 

Since $(-)^L$ and $(-)^{3B}$ remain conserved, so is the usual $R$ parity 
of the MSSM.  The neutralino mass matrix is now $9 \times 9$ instead of 
$4 \times 4$ because of the five additional neutral higgsinos 
$\tilde{\xi}^0_{1,2}$ and $\tilde{S}_{2,1,0}$.   However, only the axino 
$(u_1 \tilde{S}_1 + 2 u_0 \tilde{S}_0)/\sqrt{u_1^2+4u_0^2}$ may be light, 
and its mixing with the four usual neutralinos is very small.  The lightest 
among these five particles is a candidate for the dark matter of the 
Universe, \emph{in addition to the axion}.  Note the possibility of 
discovering a ``dark-matter'' neutralino at the LHC which is actually 
unstable but decays only into the axino with a very long lifetime.  
The exotic $h$ quarks are predicted to have masses of order $M_{SUSY}$ 
in this scenario and amenable to discovery as well.

\begin{table}[htb]
\caption{Particle content of Example 2.}
\begin{center}
\begin{tabular}{|c|c|c|c|}
\hline 
Superfield & $U(1)_{PQ}$ & 
$(-)^L$ & $(-)^{3B}$ \\ 
\hline
$Q,u^c,d^c,N$ & 1/2 & + & -- \\ 
\hline
$L$ & 0 & -- & + \\ 
$e^c$ & 1 & -- & + \\ 
\hline
$\Phi_{1,2}$ & --1 & + & + \\ 
\hline
$h,h^c$ & --1 & + & + \\ 
\hline
$(\Sigma^+,\Sigma^0,\Sigma^-)$ & 1 & + & + \\
\hline
$S_{2,1,0}$ & 2,--1,--2 & + & + \\ 
\hline
\end{tabular}
\end{center}
\end{table}

\newpage
\noindent {\bf (II)} As a second example, consider the fermion-triplet 
mechanism of neutrino mass in a supersymmetric context. In Table 3, the PQ 
charges of the superfields of this construction and their \emph{derived} 
$(-)^L$ and $(-)^{3B}$ values are listed.  The complete superpotential is 
given by
\begin{eqnarray}
W_2 &=& m_0 S_0 S_2 + \lambda_1 S_1 S_1 S_2 + \lambda_2 S_1 N N 
+ \lambda_3 S_0 \Sigma \Sigma + f_1 S_2 \Phi_1 \Phi_2 + f_2 S_2 h h^c 
+ f_3 Q Q h \nonumber \\ 
&+& f_4 u^c d^c h^c + f_5 h d^c N + f_d \Phi_1 Q d^c 
+ f_u \Phi_2 Q u^c + f_e \Phi_1 L e^c + f_\Sigma \Phi_2 L \Sigma.
\end{eqnarray}
It is easy to see that $(-)^L$ and $(-)^{3B}$ are automatically conserved 
in this case as well.  At the TeV energy scale, this model is effectively 
identical to that of Example 1.

\begin{table}[htb]
\caption{Particle content of Example 3.}
\begin{center}
\begin{tabular}{|c|c|c|c|}
\hline 
Superfield & $U(1)_{PQ}$ & 
$(-)^L$ & $(-)^{3B}$ \\ 
\hline
$Q,u^c,d^c,N$ & 1/2 & + & -- \\ 
\hline
$L$ & 3/2 (--3/2) & -- & + \\ 
$e^c$ & --1/2 (5/2) & -- & + \\ 
\hline
$\Phi_{1,2}$ & --1 & + & + \\ 
\hline
$h,h^c$ & --1 & + & + \\ 
\hline
$(\Sigma^+,\Sigma^0,\Sigma^-)$ & 1 & -- & -- \\
\hline
$S_{2,1,0}$ & 2,--1,--2 & + & + \\ 
\hline
$(\eta_1^0,\eta_1^-)$ & 1/2 (--5/2) & + & -- \\ 
$(\eta_2^+,\eta_2^0)$ & --5/2 (1/2) & + & -- \\ 
\hline
\end{tabular}
\end{center}
\end{table}

\newpage
\noindent {\bf (III)} The third example has to do with the recent development 
of using a second scalar doublet $(\eta^+,\eta^0)$ which is odd under an 
assumed $Z_2$ \cite{dm78} as a dark-matter candidate 
\cite{m06_1,bhr06,lnot07,glbe07,cmr07}. The origin of this mysterious $Z_2$ 
becomes clear if Example 2 is extended to include the superfields 
$(\eta^0_1,\eta^-_1)$ and $(\eta^+_2,\eta^0_2)$ as shown in Table 4.  
The complete superpotential is given by
\begin{eqnarray}
W_3 &=& m_0 S_0 S_2 + \lambda_1 S_1 S_1 S_2 + \lambda_2 S_1 N N + \lambda_3 S_0 
\Sigma \Sigma + f_1 S_2 \Phi_1 \Phi_2 + f_2 S_2 h h^c \nonumber \\ 
&+& f_3 Q Q h + f_4 u^c d^c h^c + f_5 h d^c N + f_d \Phi_1 Q d^c 
+ f_u \Phi_2 Q u^c + f_e \Phi_1 L e^c \nonumber \\ 
&+& f_\eta S_2 \eta_1 \eta_2 + f_\Sigma \eta_2 L \Sigma + f_N \eta_1 \Phi_2 N 
~(f_N \eta_2 \Phi_1 N).
\end{eqnarray}
Note that $(L,e^c,\eta_1,\eta_2)$ have two possible sets of PQ values: either 
$(3/2,-1/2,1/2,-5/2)$ for the choice $\eta_1 \Phi_2 N$ or 
$(-3/2,5/2,-5/2,1/2)$ in the case of $\eta_2 \Phi_1 N$.  Again the $(-)^L$ 
and $(-)^{3B}$ values are derived from $U(1)_{PQ}$ and their conservation 
implies $R$ parity conservation of the MSSM.  The scalar $\eta_{1,2}$ 
doublets are predicted to be at the TeV scale, but they have odd $(-)^{3B}$, 
i.e. odd $R$ parity, and may thus be considered as dark-matter candidates. 
They are connected to the usual MSSM particles of odd $R$ parity through 
either $N$ (for higgsinos) or $\Sigma$ (for scalar leptons).  Since both 
$N$ and $\Sigma$ are very heavy, $\eta_{1,2}$ belong effectively to a separate 
class of dark-matter candidates, in analogy to the case of the axino discussed 
earlier.  In this scenario, cosmological dark matter is composed of the 
axion and one of three particles of odd $R$ parity: the axino, the lightest 
MSSM neutralino, and the lightest neutral scalar contained in $\eta_{1,2}$. 
At the LHC, the axino is not likely to be discovered because it has very 
small couplings to ordinary matter, but the other two dark-matter candidates 
may both appear as missing energy, even if neither is the true cosmological 
dark matter. 

Because of the conserved $Z_2 = (-)^{3B}$, neutrino masses are generated 
radiatively \cite{m06_1,m06_2,m06_3,hkmr07,ms07,m08_3,m08_4,m08_5} as shown 
in Fig.~1.  The left or right diagram corresponds to choosing
$\eta_1 \Phi_2 N$ or $\eta_2 \Phi_1 N$ in $W_3$ of Eq.~(7) respectively.

\begin{figure}[htb]
\begin{center}\begin{picture}(500,100)(10,45)
\ArrowLine(70,50)(110,50)
\ArrowLine(150,50)(190,50)
\ArrowLine(150,50)(110,50)
\ArrowLine(230,50)(190,50)
\Text(90,35)[b]{$\nu$}
\Text(210,35)[b]{$\nu$}
\Text(152,35)[b]{$\Sigma^0$}
\Text(152,95)[b]{$N$}
\Text(102,60)[b]{$\eta^0_2$}
\Text(111,78)[b]{$\eta^0_1$}
\Text(199,61)[b]{$\eta^0_2$}
\Text(188,78)[b]{$\eta^0_1$}
\Text(110,116)[b]{$\phi^0_2$}
\Text(190,116)[b]{$\phi^0_2$}
\DashArrowLine(115,111)(130,85){3}
\DashArrowLine(185,111)(170,85){3}
\DashArrowArc(150,50)(40,140,180){3}
\DashArrowArcn(150,50)(40,160,120){3}
\DashArrowArc(150,50)(40,60,100){3}
\DashArrowArcn(150,50)(40,40,0){3}
\DashArrowArc(150,50)(40,20,60){3}
\DashArrowArcn(150,50)(40,120,80){3}

\ArrowLine(270,50)(310,50)
\ArrowLine(350,50)(390,50)
\ArrowLine(350,50)(310,50)
\ArrowLine(430,50)(390,50)
\Text(290,35)[b]{$\nu$}
\Text(410,35)[b]{$\nu$}
\Text(352,33)[b]{$\Sigma^0$}
\Text(352,95)[b]{$N$}
\Text(305,70)[b]{$\eta^0_2$}
\Text(396,70)[b]{$\eta^0_2$}
\Text(310,116)[b]{$\phi^0_1$}
\Text(390,116)[b]{$\phi^0_1$}
\DashArrowLine(330,85)(315,111){3}
\DashArrowLine(370,85)(385,111){3}
\DashArrowArc(350,50)(40,120,180){3}
\DashArrowArc(350,50)(40,60,100){3}
\DashArrowArcn(350,50)(40,60,0){3}
\DashArrowArcn(350,50)(40,120,80){3}

\end{picture}
\end{center}
\caption[]{One-loop radiative contributions to neutrino mass.}
\end{figure}
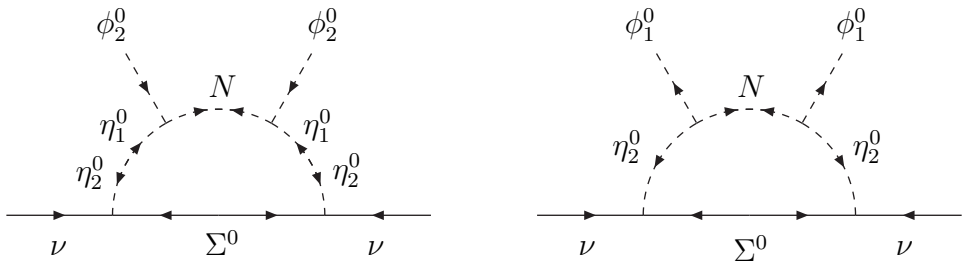

In conclusion, it has been shown how the benchmark theoretical framework 
of the Supersymmetric Standard Model can be improved.  The key is to take 
care of the strong CP problem using $U(1)_{PQ}$.  Two mass scales emerge, 
one corresponding to the axion scale $f_a$, the other the soft supersymmetry 
breaking scale $M_{SUSY}$.  With the appropriate particle content and 
$U(1)_{PQ}$ assignment, the former is identified with the seesaw scale of 
neutrino mass generation, breaking additive lepton number $L$ to 
multiplicative lepton number $(-)^L$ as well as separately baryon number 
$B \to (-)^{3B}$, the latter is identified with particles at the TeV scale 
as well as the seed of electroweak symmetry breaking.  Unconstrained 
baryogenesis is implemented by the decay of a neutral electroweak singlet 
$N$ into final states of opposite baryon number.  Exotic quarks of charge 
$\mp 1/3$ and $B = \mp 2/3$ are predicted at the TeV scale.  Neutrino mass 
is obtained in three examples: using (I) heavy electroweak triplets 
$(\xi^{++}_1,\xi^+_1,\xi^0_1)$ and $(\xi^0_2,\xi^-_2,\xi^{--}_2)$, (II) heavy 
electroweak triplets $(\Sigma^+,\Sigma^0,\Sigma^0)$, and (III) light 
electroweak doublets $(\eta^0_1,\eta^-_1)$ and $(\eta^+_2,\eta^0_2)$ in 
additional to the heavy $(\Sigma^+,\Sigma^0,\Sigma^0)$.  In (III), neutrino 
masses are radiatively generated and the extra $Z_2$ assumed in recent 
proposals for electroweak doublet scalar dark matter (DSDM) is identified 
with $(-)^{3B}$.  The true cosmological dark matter is the axion (which has 
even $R$ parity and decays into two photons), together with the DSDM or the 
LSP in the MSSM or the axino, whichever of the three is the lightest.  
However, the effective interactions connecting these three dark-matter 
candidates of odd $R$ parity are very weak, so even though the two heavier 
ones decay, they do so very slowly.  At the LHC, they would appear 
only as missing energy.

This work was supported in part by the U.~S.~Department of Energy under 
Grant No.~DE-FG03-94ER40837.

\bibliographystyle{unsrt}

\end{document}